\documentclass[
    ,final            % use final for the camera ready runs
    ,numberedheadings % uncomment this option for numbered sections
  ]
  {aipproc}

\layoutstyle{6x9}

\begin{document}

\title{Measurements of the CKM angle $\phi_3/\gamma$}


\author{T.A.-Kh.~Aushev}{
  address={
for the Belle Collaboration\\
Swiss Federal Institute of Technology of Lausanne, Switzerland\\
Institute for Theoretical and Experimental Physics, Moscow, Russia\\
aushev@itep.ru}
}

\begin{abstract}
In this report we summarize the most recent results of measurements of
the angle $\gamma/\phi_3$ of the Unitarity Triangle.
\end{abstract}

\maketitle

\section{Introduction}

Measurements of the Unitarity Triangle parameters allow to search for
New Physics effects at low energies.  One angle, $\phi_1$ (or
$\beta$)\footnote{Two different notations of the Unitarity Triangle
  are used: $\alpha$, $\beta$, $\gamma$ or $\phi_2$, $\phi_1$ and
  $\phi_3$, respectively.}, has been measured with high precision at
BaBar and Belle experiments, the $B$ factories operated at the
$e^+e^-$ colliders with the center-of-mass energy at $\Upsilon(4S)$
resonance.  The measurements of the angle $\phi_2/\alpha$ is more
difficult due to theoretical uncertainties in the calculation of the
penguin diagram contribution.  Precise determination of the third
angle, $\phi_3/\gamma$, is possible, $e.g.$, in the decays $B^\pm\to
DK^\pm$.  Although it is theoretically clean due to the absence of
loop contributions, it requires a lot more data than for the other
angles.  This report summarizes the most recent progress in measuring
the angle $\phi_3/\gamma$.

\section{GLW analyses}

The theoretically clean measurement technique was suggested by Gronau,
London, and Wyler (GLW).  It exploits the interference between $B^-\to
D^0K^-$ and $B^-\to\bar D^0K^-$ decay amplitudes, where the $D^0$ and
$\bar D^0$ mesons decay to the same $CP$ eigenstate ($CP$-even $D_1\to
K^+K^-$, $\pi^+\pi^-$ and $CP$-odd $D_2\to K_S^0\pi^0$, $K_S\omega$,
$K_S^0\phi$...)~\cite{glw}.  The following variables are used to
extract $\phi_3$ using the GLW method: the asymmetries
\begin{equation}
{\cal A}_{1,2}\equiv
\frac
{{\cal B}(B^-\to D_{1,2}K^-)-{\cal B}(B^+\to D_{1,2}K^+)}
{{\cal B}(B^-\to D_{1,2}K^-)+{\cal B}(B^+\to D_{1,2}K^+)} = 
\frac
{2r_B\sin\delta^\prime\sin\phi_3}
{1+r_B^2+2r_B\cos\delta^\prime\cos\phi_3}
\end{equation}
and the ratios
\begin{equation}
{\cal R}_{1,2}\equiv
\frac
{{\cal B}(B^-\to D_{1,2}K^-)+{\cal B}(B^+\to D_{1,2}K^+)}
{{\cal B}(B^-\to D^0K^-)+{\cal B}(B^+\to\bar D^0K^+)} = 
1+r_B^2+2r_B\cos\delta^\prime\cos\phi_3,
\end{equation}
where $\delta^\prime=\delta_B$ for $D_1$ and $\delta_B+\pi$ for $D_2$
and $r_B\equiv|A(B^-\to\bar D^0K^-)/A(B^-\to D^0K^-)|$ is the ratio of
the $b\to c\bar us$ and $b\to u\bar cs$ magnitudes, $\delta_B$ is
their strong-phase difference.  It can also be expressed in terms of three independent quantities:
\begin{equation}
x_\pm=r_B\cos(\delta_B\pm\phi_3)=
\frac{{\cal R}_1(1+{\cal A}_1)-{\cal R}_2(1+{\cal A}_2)}{4}
{\rm\ \ \ \ and\ \ \ \ \ }
r_B^2=\frac{{\cal R}_1+{\cal R}_2-2}{2}.
\end{equation}

Recently, BaBar updated their GLW analysis using the data sample of
382M $B\bar B$ pairs~\cite{glw_babar}.  These results are presented in
Table~\ref{glw_table}.  Also results from Belle using 275M $B\bar B$
pairs are shown~\cite{glw_belle}.

\begin{table}
\begin{tabular}{lrr}
\hline
& BaBar & Belle \\
\hline
${\cal R}_1$ & 
$1.06\pm0.10\pm0.05$ &
$1.13\pm0.16\pm0.05$ \\
${\cal R}_2$ & 
$1.03\pm0.10\pm0.05$ &
$1.17\pm0.14\pm0.14$ \\
${\cal A}_1$ & 
$0.27\pm0.09\pm0.04$ &
$0.06\pm0.14\pm0.05$ \\
${\cal A}_2$ & 
$-0.09\pm0.09\pm0.02$ &
$-0.12\pm0.14\pm0.05$ \\
$x_+$ &
$-0.09\pm0.05\pm0.02$ &
$-0.06\pm0.08\pm0.05$ \\
$x_-$ &
$0.10\pm0.05\pm0.03$ &
$0.04\pm0.08\pm0.04$ \\
$r_B^2$ &
$0.05\pm0.07\pm0.03$ &
$0.15\pm0.11\pm0.08$ \\
\hline
\end{tabular}
\caption{Results of the GLW analyses}
\label{glw_table}
\end{table}

\section{ADS method}

The difficulties in the application of the GLW methods arise primarily
due to the small magnitude of the $CP$ asymmetry of the $B^+\to
D_{CP}K^+$ decay probabilities, which may lead to significant
systematic uncertainties in the observation of the $CP$ violation.  An
alternative approach was proposed by Atwood, Dunietz and
Soni~\cite{ads}.  Instead of using the $D^0$ decays to $CP$
eigenstates, the ADS method uses Cabibbo-favored and doubly
Cabibbo-suppressed decays: $\bar D^\to K^-\pi^+$ and $D^0\to
K^-\pi^+$.  In the decays $B^+\to\big[K^-\pi^+\big]_DK^+$ and
$B^-\to\big[K^+\pi^-\big]_DK^-$, the suppressed $B$ decay corresponds to the
Cabibbo-allowed $D^0$ decay, and vice versa.  Therefore, the
interfering amplitudes are of similar magnitudes, and one can expect
the significant $CP$ asymmetry.

Unfortunately, the branching ratios of the decays mentioned above are
so small that they cannot be observed using the current experimental
statistics.  The observable that is measured in the ADS method is the
fraction of the suppressed and allowed branching ratios:
\begin{equation}
{\cal R}_{ADS}=
\frac
{{\cal B}(B^\pm\to\big[K^\mp\pi^\pm\big]_DK^\pm)}
{{\cal B}(B^\pm\to\big[K^\pm\pi^\mp\big]_DK^\pm)}
=r_B^2+r_D^2+2r_Br_D\cos\phi_3\cos\delta,
\end{equation}
where $r_D$ is the ratio of the doubly Cabibbo-suppressed and
Cabibbo-allowed $D^0$ decay amplitudes:
\begin{equation}
r_D=\big|\frac{A(D^0\to K^+\pi^-)}{A(D^0\to K^-\pi^+)}\big|=0.0578\pm0.0008,
\end{equation}
and $\delta$ is a sum of strong phase differences in $B$ and $D$
decays: $\delta=\delta_B+\delta_D$.

The update of the ADS analysis using 657M $B\bar B$ pairs is reported
by Belle~\cite{ads_belle}.  In the absence of the signal the ratio of
the suppressed and allowed modes is measured to be: ${\cal
  R}_{ADS}<1.8\times10^{-2}$ at 90\% C.L., which corresponds to
$r_B<0.19$.

Recently, the ADS analysis has been performed by BaBar using $B^0\to
D^0K^{*0}$ decay modes with $D^0\to K^-\pi^+, K^-\pi^+\pi^0$ or
$K^-\pi^+\pi^+\pi^-$, based on 465M $B\bar B$ events~\cite{ads_babar}.
Neglecting $K^{*0}$ final state interference and combining three $D^0$
decay modes the 95\% probability range for $r_S$ is found to be:
\begin{equation}
{\cal R}_{ADS}\approx r_s^2=\frac
{\Gamma(B^0\to D^0K^+\pi^-)}
{\Gamma(B^0\to\bar D^0K^+\pi^-)}=[0.07,0.41].
\end{equation}

\section{Dalitz plot analyses}

A Dalitz plot analysis of a $D$ meson three-body final state allows to
obtain all the information required for a $\phi_3$ determination.
Three-body final states such as $K_S^0\pi^+\pi^-$~\cite{dalitz,
  dalitz_bondar} have been suggested as promising modes for the
extraction of $\phi_3$.  Like GLW or ADS method, the two amplitudes
interfere as the $D^0$ and $\bar D^0$ mesons decay into the same final
state $K_S^0\pi^+\pi^-$.  Once the $D^0\to K_S^0\pi^+\pi^-$ decay
amplitude, $f_D$, is known, a simultaneous fit of $B^+$ and $B^-$ data
allows the contributions of $r_B$, $\phi_3$ and $\delta_B$ to be
separated.

Both Belle and BaBar collaborations reported recently the updates of
the $\phi_3/\gamma$ measurements using Dalitz plot analysis.  The
preliminary results obtained by Belle~\cite{dalitz_belle} uses the
data sample of 657M $B\bar B$ pairs and two modes, $B^\pm\to DK^\pm$
and $B^\pm\to D^*K^\pm$ with $D^*\to D\pi^0$.  The neutral $D$ meson
is reconstructed in $K_S^0\pi^+\pi^-$ final state in both cases.  The
$f_D$ is determined from a large sample of flavor-tagged $D^0\to
K_S^0\pi^+\pi^-$ decays produced in continuum $e^+e^-$ annihilation.

BaBar uses a smaller data sample of 383M $B\bar B$
pairs~\cite{dalitz_babar}, but analyses seven different decay modes:
$B^\pm\to DK^\pm$, $B^\pm\to D^*K^\pm$ with $D^*\to D\pi^0$ and
$D\gamma$, and $B^\pm\to DK^{*\pm}$, where the neutral $D$ meson is
reconstructed in $K_S^0\pi^+\pi^-$ and $K_S^0K^+K^-$ (except for
$B^\pm\to DK^{*\pm}$) final states.  The K-matrix formalism is used by
default to describe the $\pi\pi$ $S$-wave, while the $K\pi$ $S$-wave
is parametrized using $K_0^*(1430)$ resonances and as effective range
non-resonant component with a phase shift.

The results of both analyses are summarized in
Table~\ref{dalitz_table}.
\begin{table}
\begin{tabular}{lcc}
\hline
& BaBar & Belle \\
\hline
$\phi_3/\gamma$ &
$76^{+23}_{-24}\ ^{\rm o}\pm5^{\rm o}\pm5^{\rm o}$ ($3.0\sigma$ CPV) &
$76^{+12}_{-13}\ ^{\rm o}\pm4^{\rm o}\pm9^{\rm o}$ ($3.5\sigma$ CPV) \\
$r_B$   & $0.086\pm0.035\pm0.010\pm0.011$ & $0.16\pm0.04\pm0.01\pm0.05$ \\
$r_B^*$ & $0.135\pm0.051\pm0.011\pm0.005$ & $0.21\pm0.08\pm0.01\pm0.05$ \\
\hline
\end{tabular}
\caption{Results of the Dalitz plot analyses}
\label{dalitz_table}
\end{table}

\section{$B^0\to D^{*-}\pi^+$ analyses}

The study of the time-dependent decay rates of $B^0(\bar B^0)\to
D^{*\mp}\pi^\pm$ provides a theoretically clean method for extracting
$\sin(2\phi_1+\phi_3)$ from the interference of doubly
Cabibbo-suppressed (DCS) $b\to u$ and Cabibbo-favored (CF) $b\to c$
amplitudes.  The $CP$ violation parameters are given by
\begin{equation}
S^{\pm}=-2R\sin(2\phi_2+\phi_3\pm\delta),
\end{equation}
where $\delta$ is the strong phase difference between DCS and CF
decays.  Since the ratio $R$ of these two amplitudes is small,
$\approx0.02$~\cite{dstpi_r}, the amount of $CP$ violation is expected
to be small and a large data sample is needed in order to obtain
sufficient sensitivity.  Recently, Belle updated their result using
657M $B\bar B$ pairs data sample with a partial reconstruction
technique, wherein the signal is distinguished from background on the
basis of kinematics of the 'fast' pion ($\pi_f$) from $B\to D^*\pi_f$,
and the 'soft' pion ($\pi_s$) from the subsequent decay of $D^*\to
D\pi_s$; the $D$ meson is not reconstructed at all~\cite{dstpi_belle}.
To suppressed the increased background a high momentum lepton is
required in the event.  This lepton also provides the information
about the flavor and decay vertex of the tagging $B$ meson, which is
necessary for the time-dependent analysis.  The vertex of the signal
$B$ is reconstructed using 'fast' pion only.

The fit results using the partial reconstruction method are:
\[S^+=+0.057\pm0.019\pm0.012,\]
\[S^-=+0.038\pm0.020\pm0.010,\]
with $2.6\sigma$ significance of a non-zero $CP$ violation.  Together
with other measurements the significance of $CP$ violation in $B\to
D^{(*)}h$ decays is $\approx4.0\sigma$.

\section{Conclusion}

In the past year, many new measurements of angle $\phi_3/\gamma$ are
provided by Belle and BaBar.  The most precise measurements are
performed using Dalitz plot analyses of $B\to DK$ decays in a good
agreement between both experiments.

\end{document}